\documentstyle[epsfig,rotating]{aipproc}

\newcommand{\lsim}   {\mathrel{\mathop{\kern 0pt \rlap
  {\raise.2ex\hbox{$<$}}}
  \lower.9ex\hbox{\kern-.190em $\sim$}}}
\newcommand{\gsim}   {\mathrel{\mathop{\kern 0pt \rlap
  {\raise.2ex\hbox{$>$}}}
  \lower.9ex\hbox{\kern-.190em $\sim$}}}
\def\be{\begin{equation}}
\def\ee{\end{equation}}
\def\ba{\begin{eqnarray}}
\def\ea{\end{eqnarray}}
\def\ap{\approx}

\font\menge=bbold9 scaled \magstep1
\def\nota#1{\hbox{$#1\textfont1=\menge $}}
\def\R{\nota R}

\begin{document}
\title{Ultrahigh energy cosmic rays and new particle physics%
       \footnote{Invited talk at ``CAPP 2000'', Verbier, July 2000
                 \hfill {\small CERN-TH 2000-324}}}

\author{M.~Kachelrie{\ss}}
\address{TH Division, CERN, CH-1211 Geneva 23}
\maketitle

\begin{abstract}
The current status of the UltraHigh Energy Cosmic Ray (UHE CR) enigma 
and several proposed solutions involving particle physics beyond the
standard model are discussed. Emphasis is given to top--down models, and  
as a main example, supermassive dark matter as galactic source for
UHE CR and the status of its experimental signatures (galactic
anisotropy, chemical composition and clustering) is reviewed. Then
different approaches to calculate fragmentation spectra of
supermassive particles are discussed. Finally, it is argued 
that UHE neutrinos cannot be -- neither directly or indirectly -- 
responsible for the observed vertical air showers.
\end{abstract}

%%%%%%%%%%%%%%%%%%%%%%%%%%%%%%%%%%%%%%%%%%%%%%%%%%%%%%%%%%%%%%%%%%%
\section{Introduction}

Cosmic Rays~\cite{CR} (CR) are observed in a wide energy range,
starting from subGeV energies up to $3\times 10^{20}$~eV. Apart from
the highest energies, these particles are accelerated in our Galaxy,
most probably by shocks produced by SNII explosions.
Since the galactic magnetic field cannot confine and isotropize
particles with energy higher than $\sim Z\times 10^{19}$~eV, it is natural
to think that the UltraHigh Energy (UHE) component has an
extragalactic origin.  
Moreover, the acceleration of protons or nuclei up to 2--3$\times 10^{20}$~eV
is difficult to explain within the known astrophysical galactic sources. 

The most prominent signature of extragalactic UHECR is the so called
Greisen--Zatsepin--Kuzmin (GZK) cutoff~\cite{GZK}: 
the energy losses of protons 
sharply increase at $E_{\rm GZK} \ap 6\times 10^{19}$~eV, owing to
scattering on photons from the microwave background,
$p+\gamma_{3K}\to\Delta^+ \to N+\pi$, reducing their
mean free path length to less than  
$50$~Mpc or so. Nuclei exhibit a similar cutoff at the same energy, while
photons have no sharp cutoff but an even shorter free mean path,
because of
pair-production on the radio background. Thus, the UHECR spectrum
should dramatically steepen above $E_{\rm GZK}$ for {\em any\/}
homogeneous distribution of proton or nuclei sources. However,
the observed spectrum extends up to $3\times 10^{20}$~eV and this
maximum seems to be caused by the limited exposure. Both spectra,
the expected and the observed one, are shown in Fig.~1, taken from
Ref.~\cite{agasa00}.

There is another argument that disfavours the standard
astrophysical sources: At energies $E\sim 10^{20}$~eV, the arrival
direction of the primaries which is known within several
degrees should point towards their site of origin. But 
no source of UHECR such us active galactic nuclei (AGN) has been
found within $50$~Mpc in the direction of these events.
In Fig.~\ref{cluster}, the arrival directions of 92 events above
$4\times 10^{19}$~eV are shown in
galactic coordinates~\cite{Uc00}. The events are
scattered isotropically on larger scales, and no significant
enhancement towards the galactic or supergalactic plane is found. 
Intriguingly, about 20\% of the events are clustered in angular doublets
or even triplets; both triplets are found near the supergalactic plane. 

\begin{figure}
\begin{center}
\epsfig{file=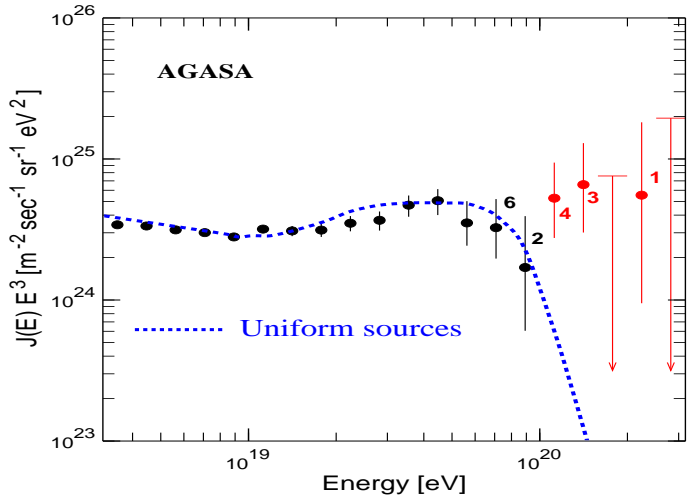, height=7cm,width=10cm}
\caption{Energy spectrum as observed by AGASA. Error
  bars are 68\% C.L., numbers are the number of events per energy bin.
  The dashed line is the spectrum expected from uniformly
  distributed astrophysical sources (from Ref.~[3]).
\label{agasa}}
\end{center}
\end{figure}

An elegant solution to some of the problems described above are
top--down models: in 
contrast to the standard sources, the primaries are not accelerated but
are the fragmentation products of some decaying superheavy particle
$X$. For $X$-particles with mass $m_X\gsim 10^{12}$~GeV,
the acceleration problem is solved trivially. Moreover, 
these sources also evade detection by normal astronomical methods.

This review is organised as follows:
In Sec.~II top-down models and in Sec.~III their signatures are presented.
We then discuss how the spectrum of hadrons produced in the decay of
supermassive particles can be calculated. 
Finally, we argue  in Sec.~5 that UHE neutrinos cannot be -- neither
directly or indirectly via resonant annihilations on relic neutrinos
-- responsible for the observed UHECR events.

For a review of astrophysical sources, we refer the
interested reader to Ref.~\cite{astro}. The suggestion that the
violation of Lorentz invariance induced by quantum gravity
effects avoids the GZK cutoff is covered in the recent review~\cite{El00}.

\begin{figure}
\begin{center}
\epsfig{file=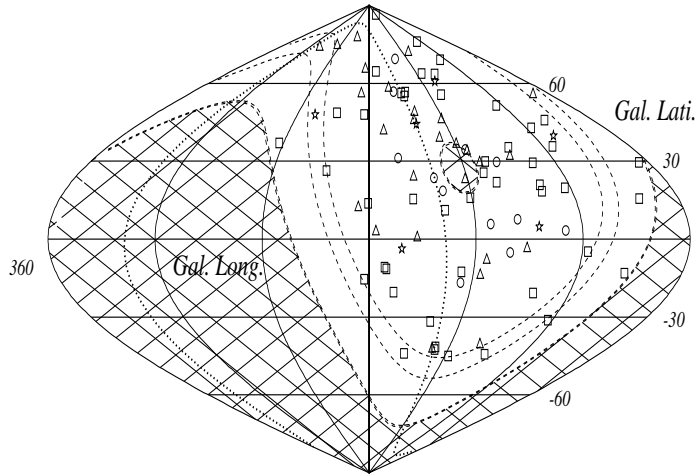, height=7cm,width=10cm}
\caption{\label{cluster}
Map of 92 events above $4\times 10^{19}$~eV in galactic
  coordinates (from Ref.~[4]). The supergalactic plane is
  shown by the dotted line, the hatched region is unobservable.}
\end{center}
\end{figure}

%%%%%%%%%%%%%%%%%%%%%%%%%%%%%%%%%%%%%%%%%%%%%%%%%%%%%%%%%%%%%%%%%%%%%
\section{Top--down models}
Top--down model is a generic name for all proposals in which the
observed UHECR primaries are produced as decay products of some
superheavy particles $X$. These $X$ particles can be either metastable
or be emitted by topological defects at the present epoch.

1. {\em Topological defects\/} (TD)~\cite{td} such as (superconducting)
cosmic strings,  monopoles, and hybrid defects can be
effectively produced in non-thermal phase transitions during the
preheating stage~\cite{Kh98/Ku99}. Therefore the presence of TDs
is not in conflict with an inflationary period of the early
Universe. They can naturally produce particles with large enough
energies but fail generally to produce the observed flux of UHE
primaries: The observation of the low energy diffuse gamma radiation
by EGRET limits the energy dumped by high-energy particles into
electromagnetic cascades and thereby also severely the possible UHECR flux.   
Another general reason for the low fluxes is the large distance
between TDs, which is often comparable to the Hubble radius. Then the
flux of UHE particles is either exponentially suppressed or strongly
anisotropic if a TD is by chance nearby.
 
{\em Ordinary strings\/} can produce UHE particles e.g. when string loops
self-intersect or when two cusp segments overlap and annihilate. In
the latter case, the maximal energy of the produced fragmentation
products is not $m_X/2$ but can be much larger due to the high
Lorentz factors of the ejected $X$ particles.

{\em Superconducting strings:\/}
Cosmic strings can be superconducting in a broad class
of particle models. Electric currents can be induced in
the string either by a primordial magnetic field that decreases during
the expansion of the Universe or when the string moves through
galactic fields at present. If the Fermi momentum of the the trapped
particles exceeds their mass outside the string, they can leave the
string and decay.

{\em Monopolium M\/} -- boundstates of monopole--antimonopole pairs -- was
the first TD proposed as UHECR source~\cite{Hi83}. It clusters like
Cold Dark Matter (CDM) and is therefore an example of a SuperHeavy
Dark Matter (SHDM) particle. The galactic density of monopoles
is constrained by the Parker limit: the galactic magnetic field should
not be eliminated by the acceleration of
monopoles. Reference~\cite{Bl99} concluded that the resulting limit on
the UHECR flux produced by  Monopolium annihilation is 10 orders of
magitude too low.

{\em Cosmic necklaces\/} are hybrid defects
consisting of monopoles connected by a string. These defects are produced 
by the symmetry breaking $G\to H\times U(1) \to H\times Z_2$, where
$G$ is semi-simple.
In the first phase transition at scale $\eta_m$, monopoles are 
produced. At the second phase transition, at scale  $\eta_s<\eta_m$, each 
monopole gets attached to two strings. 
%The basic parameter for the evolution
%of necklaces is the ratio $r=m/(\mu d)$ of the monopole mass $m$ and
%the mass of the string 
%between two monopoles, $\mu d$, where $\mu \sim \eta_s^2$ is the mass 
%density of the string and $d$ the distance between two monopoles.
Strings lose their energy and can contract through gravitational radiation.
As a result, all monopoles annihilate in the end producing
$X$-particles. Reference~\cite{Be97} argued that for a reasonable range of
parameters the model predicts a UHECR flux close to the observed one.
Recently, a numerical study~\cite{Si00} of the evolution of necklaces
found that the lifetime of necklaces is generally much shorter than
the age of the Universe $t_0$. A possible exception is the case
$\eta_m\gg\eta_s\sim 100$~GeV\cite{Bl99}, not a very attractive
possibility in view of the limits on the $Z^\prime$ boson mass,
$m_{Z^\prime}\gsim 600$~GeV.

2. {\em Superheavy metastable relic particles\/} (SHDM) were proposed in
Refs.~\cite{bkv97,kr97} as UHECR source. They constitute
(part of) the CDM and, consequently, their abundance in
the galactic halo is enhanced by a factor $\sim 5\times 10^4$ above
their extragalactic abundance. 
Therefore, the proton and photon flux is dominated by the halo
component and the GZK--cutoff is avoided, as was pointed out
in Ref.~\cite{bkv97}. The quotient $r_X=\Omega_X (t_0/\tau_X)$ of relic
abundance $\Omega_X$ and lifetime $\tau_X$ of the $X$-particle is
fixed by the UHECR flux, $r_X\sim {\cal O}(10^{-11})$.

The relative abundance of particles that were in thermal equilibrium 
and then freeze-out is $\Omega h^2 \propto 1/(\sigma_{\rm ann}v)$. 
Assuming for the annihilation cross-section $\sigma_{\rm ann}\lsim
1/m_X^2$ one obtains the bound $m_X\lsim 1$~TeV. 
Therefore SHDM particles should never have been in thermal
equilibrium and the challenge is to create create only few of them in
a natural way. 

There exist several plausible non-equilibrium production mechanisms. 
The most promising one is the gravitational production of the $X$
particles by the non-adiabatic change of the scale factor at the
end of inflation%, during the transition from the
%de-Sitter to the radiation dominated phase
~\cite{ch/kz}. 
In this scenario, the gravitational
coupling of the $X$-field to the background metric yields the present
abundance $\Omega_0\sim 1$ for $M_X\sim 10^{13}$~GeV, independent of
any specific particle physics model.  
Other mechanisms proposed are thermal production during reheating, 
production through inflaton decay at the preheating phase, or through
the decay of hybrid defects.

The lifetime of the superheavy particle has to be in the range
$10^{17}~{\rm s}\lsim \tau_X\lsim 10^{28}$~s, i.e. longer or much
longer than the age of the Universe. Therefore it is an obvious
question to ask if such an extremely small decay rate can be obtained
without fine-tuning. A well-known example of how metastability can be
achieved is the proton: in the standard model B--L is a conserved
global symmetry, 
and the proton can decay only via non-renormalizable operators. 
Similarly, the $X$-particle could be protected by a new global symmetry
which is only broken by higher-dimension operators suppressed by
$M^{d}$, where for instance $M\sim M_{\rm Pl}$ and $d\geq 7$ is possible. 
The case of discrete gauged symmetries has been studied in detail in
Refs.~\cite{Ha98}. Another possibility is that the global symmetry is
broken only non-perturbatively, either by wormhole \cite{bkv97} or
instanton \cite{kr97} effects. Then an exponential suppression of the
decay process is expected and lifetimes $\tau_X\gg t_0$ can be
naturally achieved. 

An example of a SHDM particle in a realistic
particle physics model is the crypton~\cite{El90}.
Cryptons are boundstates from a strongly interacting hidden sector of
string/M theory. Their mass is determined by the non-perturbative
dynamics of this sector. For example, flipped SU(5) motivated by
string theory contains boundstates with mass $\sim 10^{12}$~GeV and 
$\tau\sim 10^{15}$~yr~\cite{Be99}.

%%%%%%%%%%%%%%%%%%%%%%%%%%%%%%%%%%%%%%%%%%%%%%%%%%%%%%%%%%%%%%%%%%%%%
\section{The signatures of top-down models}
Superheavy dark matter  has three clear signatures: 
1. No GZK-cutoff, instead a flat spectrum (compared to astrophysical
sources) up to $m_X/2$. 
2. High neutrino and photon fluxes compared to the proton flux.
3. Galactic anisotropy. 
Possibly, the observed small-scale clustering gives additional
constraints.  

{\em Chemical composition:\/}
Since at the end of the QCD cascade quarks combine more easily to mesons than
to baryons, the main component of the UHE flux are neutrinos and
photons from pion decay with only a small admixture ($\sim 5\%$) of
nucleons (cf. also Fig.~\ref{MC}).   
The differentiation between photon-- and proton--induced air showers 
is however rather difficult at the highest energies:
the muon content of photon-induced showers becomes at UHE
similar to that of proton-induced showers~\cite{Ah91}. The
Landau--Pomeranchuk--Migdal effect reduces the electromagnetic
interactions at these energies, while the geomagnetic field makes the
processes $\gamma\to e^+e^-$ and $\gamma\to\gamma\gamma$ possible.
Reference~\cite{Av00} finds that above $4\times 10^{19}$~eV less than
55\% of the UHE primaries can be photons.
Since the AGASA data suggest that ``normal'' sources dominate the flux
up to $10^{20}$~eV and the flux is steeply falling with energy, this
results seems to be still not problematic for SHDM models.
Recently, Ref.~\cite{An00} claimed that the Flye's Eye event at
$3\times 10^{20}$~eV is inconsistent with a proton or photon as
primary. They compared the event with the {\em average\/}
depths from simulations instead of either only comparing the shape of
the showers or taking into account the fluctuations of the first
interaction point. Thus their conclusions, that do not agree with
those of Ref.~\cite{Ka/Ha}, seem to be artificially strong.
Future fluorescence light detectors like HiRes~\cite{HiRes} which are
able to measure the complete shower development will reliably
distinguish between photon and proton as primaries.

{\em Galactic anisotropy:\/} 
Dubovsky and Tinyakov noted that the UHECR flux from SHDM should show
a galactic anisotropy, because the Sun is not in the centre of the
Galaxy~\cite{Du98}. The degree of this anisotropy depends on
how strong the CDM is concentrated near the galactic centre -- a
question currently under debate. Since the galactic center
cannot be observed by present detectors, the predicted anisotropy is
rather small. Reference~\cite{BeMe} performed a harmonic analysis of the
UHECR flux predicted by the SHDM model and found reasonable agreement
with the AGASA data above $4\times 10^{19}$~eV, although the
statistical significance of these results is small. Similar results
were obtained in Ref.~\cite{Wi00}. We thus have to wait for the AUGER
detector~\cite{Auger} which is currently built in Argentina and can see the
galactic centre for a definite answer.

{\em Clustering:\/}
Although the distribution of UHECR arrival directions is consistent
with isotropy on larger scales, it shows an enhanced rate of clustering:
The set of 92 UHECR with $E\geq 4\times 10^{19}$~eV contains 
2 triplets and 7 doublets; above $E\geq 10^{20}$~eV, there are 2 doublets
within 14 UHECR~\cite{Uc00}. The chance probability to observe the
clustered events in the case of an isotropic distribution of arrival
directions was estimated to be $<1\%$.  

Waxman, Fisher and Piran pointed out \cite{Wa97} that the number
density $n$ of {\em uniformly\/} distributed UHECR sources can be
strongly constrained by the fraction of clustered events. As $n$
decreases, the sources have to become brighter for a fixed UHECR flux
and therefore the probability for clustering increases. The analysis
of Ref.~\cite{Du00} showed that $\sim 400$ sources should be inside
the GZK volume, compared to $\sim 10$ GRB sources or $\sim 250$ AGNs
of which only a small fraction is thought to be UHECR sources. However,
the statistical uncertainties are still very large.

The clustering probability was discussed for SHDM in
Ref.~\cite{Bl00a}. The authors assumed a smooth NFW profile
($n(r)\propto r^{-1}$ for $r\to 0$) for
the SHDM plus a clumped component and obtained a rather good 
agreement with the observed clustering statistics. Table~1 shows the
experimentally observed number of doublets $N_2$ within
$3^\circ$, $4^\circ$ and 
$5^\circ$ compared with the average numbers $\langle N_2 \rangle$
obtained in the simulation. Also shown is the probability $p$ to
observe $N_2$ or more doublets.
\begin{table}[h!]
\caption{}
\begin{tabular*}{3in}{ll@{\hspace*{3em}}|c|c|c|c|}
                      & $3^\circ$  & $4^\circ$  &   $5^\circ$ \\ 
        $N_2$           & 12       & 14       & 20        \\ 
$\langle N_2 \rangle$   & 8        & 14       & 21        \\ 
$p(\langle N_2\rangle\geq N_2)$& 12\%     & 47\%     & 57\%      
\end{tabular*}
\end{table}
Without the clumped component, the number of doublets drops typically
by 1-2, i.e. is still higher than expected from an isotropic distribution.

The signatures of TD models are not so clear.
The high photon/proton ratio at generation can be masked by
the strong absorption of UHE photons, but it is still higher than
expected from astrophysical sources.  
A crucial test can be the (non-) observation of the large
flux of UHE neutrinos predicted in all top--down models.
Finally, the GZK--cutoff is less
pronounced for TDs than for astrophysical sources, because of the
flatter generation spectrum of the UHE particles.

%%%%%%%%%%%%%%%%%%%%%%%%%%%%%%%%%%%%%%%%%%%%%%%%%%%%%%%%%%%%%%%%%%%%%
\section{Fragmentation spectrum of hadrons}

\begin{figure}
\begin{picture}(8,8.5)
 \put(2.5,1.0) {
 \epsfig{file=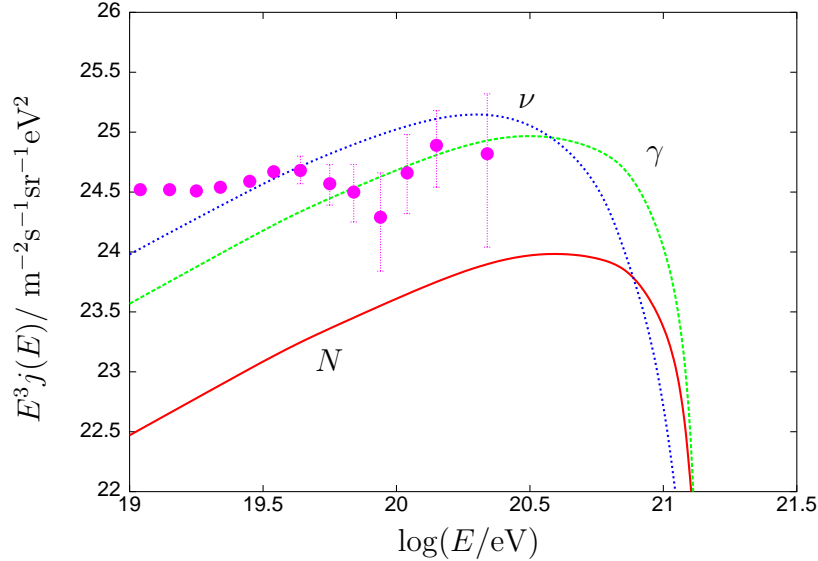, height=7cm,width=10cm} }
 \put(7.0,0.6) {$\log(E/{\rm eV})$}
 \put(1.9,2.3) {\begin{sideways}
                  $E^3j(E)/$~m$^{-2}$s$^{-1}$sr$^{-1}$eV$^2$
                \end{sideways} }
 \put(5.9,3.0) {$N$}
 \put(8.6,6.5) {$\nu$}
 \put(10.3,5.8) {$\gamma$}
\end{picture}
\caption{Flux of UHE particles from $X$-particle decay with
  $m_X=3\times 10^{12}$~GeV. The fragmentation spectra are from
  the Monte Carlo simulation~[34]; only the halo
  component is shown.
\label{MC}}
\end{figure}

To make detailed predictions about the energy spectrum of UHECR in
top-down models, the spectrum of hadrons produced in QCD cascades has
to be known. Since $m_X$ is much above the supposed mass scale 
of supersymmetric particles, SUSY partons (gluinos and
squarks) should be included in the cascade development. There are
different strategies to do this:
\begin{itemize}
\item
Analytical solutions of the DGLAP equations, in particular the
limiting spectrum, are known to describe quite accurately the observed
hadron spectra produced in $e^+e^-$ annihilations. Therefore 
the limiting spectrum for SUSY-QCD was derived and proposed as a
useful description of hadron spectra at large $m_X$ in Ref.~\cite{Be98}.

However, two assumptions are used in the derivation of the 
limiting spectrum which prevent it from being useful in the case of
SUSY-QCD: First, the number of active flavours $n_f$ has to be kept constant.
Second, the cutoff $Q_0$ for the perturbative evolutions of the
QCD cascade is identified with $\Lambda$, which fixes the strength of the
strong coupling via $\alpha_s(Q^2) = 4\pi/b\ln(Q^2/\Lambda^2)$. 
This identification makes sense for $n_f=3$, when both
parameters are $\sim 250$~MeV, but not for $n_f=6$ or even above the 
SUSY threshold.

\item
Another possibility is to evolve the fragmentation functions $D_i^a$
measured at low 
energies to the higher scale $m_X$ with the DGLAP equations~\cite{Rubin},
\be   \label{GLAP}
 \frac{\partial}{\partial\ln Q^2} D_i^a(x,Q^2)= \sum_j \int_x^1 
 \frac{dz}{z} \: \frac{\alpha_s(Q^2)}{2\pi} P_{ji}(z) D_j^a(x/z,Q^2) \,.
\ee
In this approach, the region of accessible $x$ values is restricted by
the low-energy data.

\item 
Available Monte Carlo simulations such as HERWIG or JETSET have
precision, time and memory problems for large $m_X$ and can 
be used, if at
all, only after modifications. Furthermore, supersymmetric partons are only
included as resonances at present, but not in the QCD cascade. 
In Ref.~\cite{Bi98}, HERWIG was used to calculate the spectra of
stable particles produced in $X$ decay within QCD. It was found 
that the proton flux at $x\gsim 0.2$ can compete with
the photon and neutrino flux. However, it was realized later that
the overproduction of baryons at large $x$ is an artifact of the
hadronization procedure used in HERWIG~\cite{Rubin}. 

In Ref.~\cite{Be00}, the results of a new Monte Carlo simulation,
especially written for UHE, were presented. The simulation includes
SUSY partons and a resulting spectrum for UHECR is shown in Fig.~\ref{MC}.
An advantage of this approach is that also the spectrum of the
Lightest Supersymmetric Particle (LSP) can be calculated, which is stable if
$R$-parity is conserved. The energy fraction taken away by the LSP can
be considerable~\cite{bk98} owing to its hard spectrum and was found
to be $\sim 10\%$.

\end{itemize}

\unitlength1.0cm

%%%%%%%%%%%%%%%%%%%%%%%%%%%%%%%%%%%%%%%%%%%%%%%%%%%%%%%%%%%%%%%%%%%%%
\section{UHE neutrinos}
Neutrinos are the only known stable particles that can traverse
extragalactic space without attenuation even at energies $E\gsim
E_{\rm GZK}$, thus avoiding the GZK cutoff. Therefore, it has been
speculated that the UHE primaries initiating the observed air showers
are not protons, nuclei or photons but neutrinos~\cite{alt,do98,ja00}. 
However, neutrinos are in the Standard Model (SM) deeply
penetrating particles  producing only horizontal not vertical
Extensive Air Showers (EAS).
Therefore, either one has to postulate new interaction that enhance
the UHE neutrino-nucleon cross-section or the neutrino has to be
converted locally into a strongly interacting hadron.
 
\subsection{Annihilations on relic neutrinos}
In the later scheme~\cite{We99/Fa99}, UHE neutrinos from distant sources
annihilate with relic neutrinos on the $Z$ resonance. The
fragmentation products from nearby $Z$ decays are supposed to be the
primaries responsible for the EAS above the GZK-cutoff. 
For energies of the primary neutrino of
$E_\nu\sim 10^{23}$~eV, the mass of the relic neutrino should be
$m_\nu = m_Z^2/(2E_\nu)\sim 0.1$~eV which is compatible with
atmospheric neutrino data. There are, however, severe observational
constraints on this model:

{1.} Since the Pauli principle does not allow arbitrary densely packed 
neutrinos, an upper limit for their number density $n$ 
in the galactic halo is $n\leq (4\pi/3) p_{\rm max}^3$, where 
$p_{\rm max}\sim m_\nu v_{\rm rot}$ and $v_{\rm rot}\sim 220$~km/s is
the Galactic rotation speed. A somewhat better limit comes from the
requirement that during the gravitational collapse the neutrino phase
space density does not increase. Therefore the overdensity of relic
neutrinos is small and one expects in this model a rather pronounced
GZK-cutoff.  

{2.} Since the interaction probability for a UHE neutrino in the
neutrino halo is small, a large neutrino flux is needed to produce the
observed UHECR. The limit on horizontal EAS set by the Fly's Eye
experiment~\cite{Ba85} limits therefore severely this model: Ref.~\cite{Bl00}
found that the neutrino spectrum has to be extremely flat,
$dN/dE \propto E^{-\gamma}$ with spectral index $\gamma<1.2$. Even if
one assumes a large neutrino enhancement factor due to a lepton
asymmetric Universe, the spectrum has to be much flater, $\gamma<1.8$,
than expected from astrophysical sources. 

3. The observed UHECR flux implies an upper bound on the UHE neutrino
flux produced in astrophysical sources which are not hidden. If UHE
neutrinos annihilating on relic neutrinos contribute significantly to
the observed UHECR at $\sim 10^{20}$~eV, a new class of UHE neutrino
source has to be invoked which is optically thick for nucleons.
The energy generation of these sources was estimated to be comparable
to the total photon luminosity of the Universe~\cite{Wa98}.

\subsection{UHE neutrino and weak-scale string theories}

\def\d{{\rm d}}
\def\i{{\rm i}}
\def\e{{\rm e}}
\def\ap{\approx}
\def\stot{\sigma_{\rm tot}}
\def\sel{\sigma_{\rm el}}
\def\skk{\sigma_{N\nu}^{\rm KK}}
\def\sSM{\sigma_{N\nu}^{\rm SM}}
\def\Ms{M_{\rm st}}
\def\Mp{M_{\rm Pl}}

Most models introducing new physics at a scale $M$ to produce large
cross-sections for UHE neutrinos fail because experiments generally
constrain $M$ to be larger than the weak scale, $M\gsim m_Z$, and
unitarity limits cross-sections to be ${\cal O}(\stot)\lsim 1/M^2
\lsim 1/m_Z^2$.  String theories with large extra
dimensions \cite{ex} are different in this respect: if the
SM particles are confined to the usual $3+1$-dimensional space and
only gravity propagates in the higher-dimensional space, the
compactification radius $R$ of the large extra dimensions can be
large, corresponding to a small scale $1/R$ of new physics. The
weakness of gravitational interactions is a consequence of the large
compactification radius, since Newton's constant is then given by
$G_N^{-1}=8\pi R^{\delta} M_D^{\delta+2}$, where $\delta$ is the
number of extra dimensions and $M_D\sim\;$TeV is the fundamental mass
scale.  Such a scenario is naturally realized in theories of open
strings, where SM particles correspond to open strings
beginning and ending on D-branes, whereas gravitons correspond to
closed strings which can propagate in the higher-dimensional space.
From a four-dimensional point of view the higher dimensional graviton
in these theories appears as an infinite tower of Kaluza-Klein (KK)
excitations with mass squared $m_{\vec{n}}^2=\vec{n}^2/R^2$.  Since
the weakness of the gravitational interaction is partially compensated
by the large number of KK states and cross-sections of reactions
mediated by spin 2 particles are increasing rapidly with energy, it
has been argued in Refs.~\cite{do98,ja00} that neutrinos could
initiate the observed vertical showers at the highest energies.

In the calculations of Refs.~\cite{ja00,nu99} it was assumed that the 
massless four-dimensional graviton and its massive KK excitations couple 
with the usual gravitational strength 
$\overline{M}_{\rm Pl}^{-1}= \sqrt{8\pi}/M_{\rm Pl}$. Then the sum over all KK
contributions to a given ($t$-channel) scattering amplitude,
\be
 M_{fi}\propto P(t) =
 \sum_{\vec n=1}^\infty \frac{g_{\vec n}}{t-m_{\vec n}^2} \,,
 \qquad \vec n = (n_1,\ldots,n_\delta)
\ee
only converges in the
case of one extra dimension, and for two or more extra dimensions a
cutoff has to be introduced by hand. However, it has been
pointed out \cite{fluct} that due to brane fluctuations the effective
coupling $g_{\vec{n}}$ of the level $\vec{n}$ KK mode to
four-dimensional fields is suppressed exponentially,
\begin{equation}
  g_{\vec{n}}={1\over\overline{M}_{\mbox{\small Pl}}}
  \exp\left(-{c\,m_{\vec{n}}^2\over \Ms^2}\right)\;,
  \label{coupling}
\end{equation}
where $c$ is a constant of order $1$ or larger, which parametrises the
effects of a finite brane tension \cite{fluct}, and $\Ms\sim M_D$ is the
string scale\footnote{A similar suppression of the coupling to higher
  KK modes was found also in Ref.~\cite{twist}.}.  
This exponential suppression thereby provides a
dynamical cutoff in the sum over all KK modes.

\unitlength1.0cm
\begin{figure}
\begin{picture}(8,8.5)
% \put(2.5,1.0) { \epsfig{file=sigma.eps,height=7.cm,width=10.cm,angle=0} }
 \put(2.5,1.0) {
 \epsfig{file=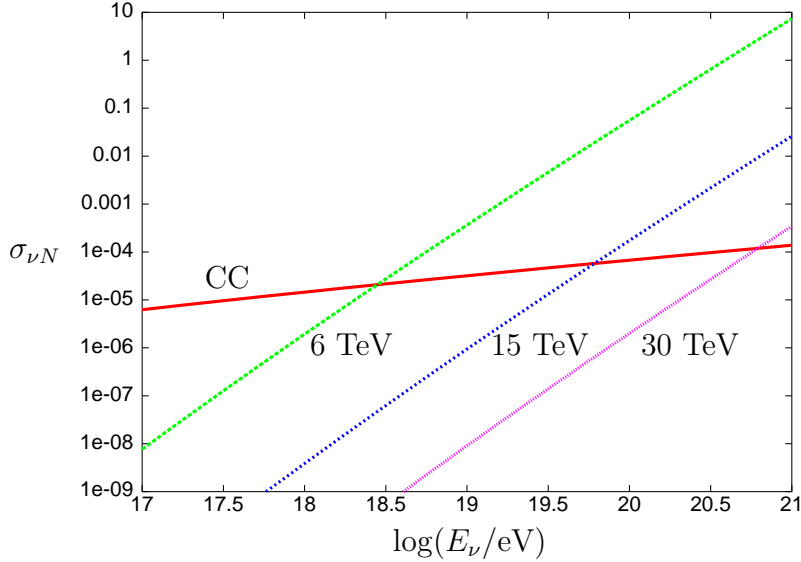,height=7.cm,width=10.cm,angle=0} } 
 \put(7.0,0.6) {$\log(E_\nu/{\rm eV})$}
 \put(1.9,4.5) {$\sigma_{\nu N}$}
 \put(4.5,4.1) {CC}
 \put(5.9,3.2) {6 TeV}
 \put(8.3,3.2) {15 TeV}
 \put(10.3,3.2) {30 TeV}
\end{picture}
\caption{Neutrino-nucleon cross-section $\sigma_{\nu N}$/mbarn due to
  $W$-exchange (CC) and exchange of KK gravitons as function
  of $\log(E_\nu/{\rm eV})$ for $\Ms=6,15$ and 30~TeV. All for
  $\delta=2$ and $c=1$.  
\label{fig_sigma}}
\end{figure}

In Fig.~\ref{fig_sigma}, the cross-sections obtained in Ref.~\cite{Ka00}
due to KK exchange are shown for three different values of $\Ms$
together with the charged-current cross-section of the SM.
It is clear that even for $\skk=10$~mbarn
a value of $\Ms$ not much above 1~TeV is required. While present collider
experiments do not exclude this possibility, SN 1987A gives $M_D\gsim 50$~TeV
\cite{bounds}. The latter limit was obtained for a rather
conservative choice of supernova parameters, and therefore
$M_D\lsim 10$~TeV seems to be incompatible with SN 1987A 
even allowing for rather large astrophysical uncertainties.

\begin{figure}
\begin{picture}(8,8.5)
% \put(2.5,1.0) { \epsfig{file=y_nu.eps,height=7.cm,width=10.cm,angle=0} }
 \put(2.5,1.0) { \epsfig{file=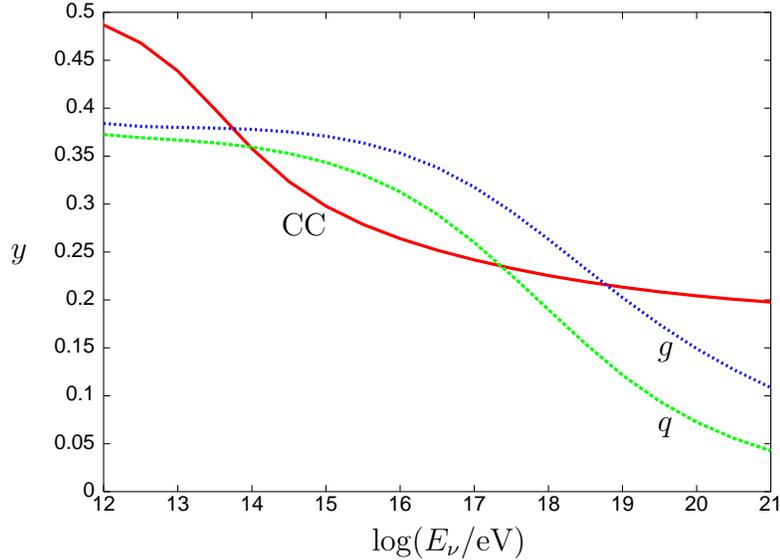,height=7.cm,width=10.cm,angle=0} }
 \put(7.0,0.6) {$\log(E_\nu/{\rm eV})$}
 \put(2.2,4.5) {$y$}
 \put(5.8,4.8) {CC}
 \put(10.8,3.2) {$g$}
 \put(10.8,2.2) {$q$}
\end{picture}
\caption{Energy transfer $y$ in the subreactions with $W$-exchange
  (CC), exchange of KK gravitons with quarks ($q$)
  and gluons ($g$) as function of $\log(E_\nu/{\rm eV})$ for
  $\Ms=6$~TeV, $\delta=2$ and $c=1$.
\label{fig_y}}
\end{figure}

The second important quantity characterising the development of an air
shower besides $\stot$  is the energy transfer 
\mbox{$y=(E_\nu-E_\nu^\prime)/E_\nu$.}  
In contrast to charged-current scattering where the
electromagnetic shower initiated by the charged lepton is practically
indistinguishable from a hadronic shower, only the hit nucleon can
initiate an air shower in KK scattering. Therefore, even a neutrino with
large $\stot$ will behave like a penetrating particle if it does not
transfer a large fraction of its energy per interaction to the
shower.
In Fig.~\ref{fig_y}, the energy transfer $y$ is shown as function of
$E_\nu$. 
At energies of interest, $E_\nu\ap 10^{20}$~eV, the transferred energy
fraction is only around $y\ap0.1$, i.e., much smaller than $y\ap 0.6$
typical of nucleon-nucleon collisions.

Let us now discuss in a very general way how large the total
cross-section of a UHE primary able to produce the observed vertical air
showers should be. The survival probability $N$ at atmospheric depth
$X$ of a primary with mean free path $\lambda=m_{\rm air}/\stot$ 
is $N(X)=\exp(-X/\lambda)$, where $m_{\rm air}\ap 2.4\cdot 10^{-24}$~g
is the weight of an ``average'' air atom. Hence, the probability
distribution $p$ of the first interaction point $X_1$ has its maximum
at $p(X_1)=\lambda$. 

For a proton with energy $E=10^{20}$~eV, the mean free path is
$\lambda_p\ap 40$g/cm$^2$ and thus a proton air shower is indeed
initiated in the top of the atmosphere. After the first interaction,
the number of particles in the shower grows until it reaches its maximum at 
$X_{\rm max}\ap 800$~g/cm$^2$. Hence, a vertical proton air shower needs
almost the complete atmosphere for its development.

How would this picture change for a neutrino with $\lambda_\nu=10\lambda_p$,
i.e., $\stot=15$~mbarn? Taking into account only the delayed start of 
the shower shifts the shower maximum already $\ap 360$~g/cm$^2$ downwards in 
the atmosphere. The small energy fraction transferred to the shower by each
interaction delays the shower development even further.
Additionally, the fluctuations of a neutrino shower are
enhanced compared to a proton shower. Hence, the shower evolution
is clearly different from a proton shower. Therefore even 
neutrino-nucleon cross-sections as large as 15~mbarn due to KK
exchange are not sufficient to explain vertical air showers by
neutrino primaries.

%%%%%%%%%%%%%%

Finally, we want to discuss the high-energy behaviour of the total
cross-section $\stot$. A partial-wave analysis shows that at
$\sqrt{s}={\cal O}(\Ms)$ the amplitude of $\nu+q\to\nu+q$ starts to
violate unitarity. At the same energy scale, one expects that 
the effective theory used to derive $\skk$ breaks down and that the
growth of $\sigma_{N\nu}^{KK}$ is slowed down. In this case, the
results shown in Fig.~\ref{fig_sigma} would be an upper bound for
$\skk$. Since a calculation of $\skk$ valid for $s\gg\Ms^2$ within string
theory seems at present not to be feasible, it is interesting to ask
if general principles uch as unitarity can be used at all as guidelines.

Khuri considered potential scattering on $\R^3\otimes S^1$ as a toy
model for string theory with one large extra dimension in
Ref.~\cite{Kh95}.  He showed that analytical properties of the
forward scattering amplitude $T_{nn}(s,t=0)$, which are true in $\R^3$,
do not necessarily hold in $\R^3\otimes S^1$ for $n>0$. At least in
this specific example however, the forward scattering amplitude for $n=0$
(``SM particles'') has the usual analytical properties known from
$\R^3$. If this would hold true generally, then the Regge approach
together with the eikonal method to ensure unitarity should give an idea 
of the high-energy behaviour of $\stot$. A general
Regge amplitude $A_R$ can be represented by
\be
 A_R(s,t)= \beta(t) s^{\alpha(t)} \,,
\ee
where the exponent $\alpha(t)$ is given by the relation between spin
$\sigma_i={\rm int}[\alpha(t)]$ and mass $m^2_i=t$ of the particles
lying on the leading Regge trajectory contributing to the
reaction. In our case, the intercept $\alpha(0)$ of this 
trajectory is equal to the spin of the massless graviton, $\alpha(0)=2$. 
String theory suggests that the Regge trajectories 
are linear, $\alpha(t)=\alpha_0+\alpha't$, and that their slope is
given by the string tension, $\alpha'=1/(4\pi\Ms^2)$. The residue
\be 
 \beta(t)= -\exp(-\i\alpha(t)\pi/2) \: \e^{at} 
\ee
contains the phase of the amplitude and the Reggeon coupling 
$\propto\exp(at)$, for which Eq.~(\ref{coupling}) gives
$a=c/\Ms^2$. In $d=4$ dimensions, the energy dependence of the
total cross-section follows as $\stot(s)\propto \ln^{2}(s)$~\cite{Ka00}.
The results of Ref.~\cite{Ch88/Pe00} suggest that for $s\gg \Ms^2$ 
and $\delta$ extra dimensions, 
$\ln^{2}(s/s_0)$ should be replaced by $\ln^{2+\delta}(s/s_0)$.

%%%%%%%%%%%%%%%%%%%%%%%%%%%%%%%%%%%%%%%%%%%%%%%%%%%%%%%%%%%
\section{Conclusions}
Superheavy, metastable relic particles is the most promising 
source in the framework of top--down models. Its cleanest signature is
the galactic anisotropy which should be easily detectable by a
detector in the southern 
hemisphere as AUGER. Experimentally more difficult signatures are  
the high photon/proton ratio and the detection of the predicted large
UHE neutrino fluxes via horizontal EAS. All top--down models predict
also an appreciable LSP flux if R-parity is conserved.

Also in theories with large extra dimensions, neutrinos behave 
as deeply penetrating particles and are therefore not responsible for the
observed UHECR events.

%%%%%%%%%%%%%%%%%%%%%%%%%%%%%%%%%%%%%%%%%%%%%%%%%%%%%%%%%%%
\section*{Acknowledgements}
I am grateful to V.~Berezinsky, S.~Ostapchenko and M.~Pl\"umacher 
for very pleasant collaborations and useful discussions.

\end{document}